\documentclass[twocolumn,pra,showpacs,superscriptaddress,floatfix]{revtex4}
\usepackage{amsmath}
\usepackage{graphicx}
\usepackage{amssymb}
\usepackage{amsmath}

\newcommand{\e}{{\textrm e}}
\newcommand{\I}{{\textrm i}}
\newcommand{\NOT}{\textsc{not}}
\newcommand{\Fn}[1]{\ensuremath{\textrm{F}_{#1}}}
\newcommand{\tdeg}[1]{\ensuremath{{#1}^\circ}}

\begin{document}
\title{Designing short robust NOT gates for quantum computation}
\author{Jonathan A. Jones}\email{jonathan.jones@qubit.org}
\affiliation{Centre for Quantum Computation, Clarendon Laboratory, University of Oxford, Parks Road, OX1 3PU, United Kingdom}
\date{\today}
\begin{abstract}
Composite pulses, originally developed in Nuclear Magnetic Resonance (NMR), have found widespread use in experimental quantum information processing (QIP) to reduce the effects of systematic errors. Most pulses used so far have simply been adapted from existing NMR designs, and while techniques have been developed for designing composite pulses with arbitrary precision the results have been quite complicated and have found little application.  Here I describe techniques for designing short but effective composite pulses to implement robust \NOT\ gates, bringing together existing insights from NMR and QIP, and present some novel composite pulses.
\end{abstract}
\maketitle

Quantum Information Processing (QIP) is the encoding of information in two level quantum systems called qubits, and the manipulation of this information through a series of unitary transformations which can be interpreted as logic gates \cite{Bennett2000}.  Building real quantum computers will require the ability to perform accurate unitary transformations on quantum systems in the presence of realistic errors.  Such errors can be divided into two broad categories: random errors, arising from decoherence, which can be tackled by methods such as quantum error correction \cite{Shor1995,Steane1996}, and systematic errors, arising from imperfections in control fields.  If systematic errors vary slowly in time, or if they vary over a spatial ensemble of qubits, it is necessary to design control sequences which are intrinsically tolerant of a range of error values.

One particularly successful approach, adopted from Nuclear Magnetic Resonance (NMR) experiments, is the use of composite pulses \cite{Levitt1986,Cummins2003,Jones2011,Merrill2012}, in which a single rotation about some axis in the $xy$-plane is replaced by a sequence of such rotations, such that the combined propagator implements the desired rotation in the absence of errors, while in the presence of small errors the errors in individual rotations do not accumulate but instead mostly cancel one another.  Many traditional composite pulses widely used in NMR are not suitable for QIP, as they are ``point to point'' pulses which make assumptions about the initial and final states.  For example, composite inversion pulses \cite{Levitt1979,Tycko1984,Tycko1985a} are designed to interconvert the computational basis states, but do not act correctly on superposition states.  However so-called Class A composite pulses, or general rotors, which are error tolerant for any initial state, \textit{are} suitable \cite{Levitt1986,Cummins2003,Jones2011}.

Here I will concentrate on attempts to construct \NOT\ gates, that is $\pi$ rotations about the $x$-axis of the Bloch sphere, using only $\pi$ rotations around axes in the $xy$-plane.  Composite pulses of this kind have many desirable properties, which will allow design techniques to be explored while sidestepping many complexities.  Although \NOT\ gates find only limited application in theoretical QIP they play an important role in many experimental techniques, such as dynamical decoupling \cite{Viola1999,Souza2011b,Souza2012}.  As with other composite pulses developed in NMR these can be applied in a wide range of other experiments \cite{Gulde2003,Collin2004,Morton2005a,Clayden2012,Ivanov2012}.

I will describe each rotation (pulse) by its propagator
\begin{equation}
\theta_\phi=\cos(\theta/2)\openone-\I\sin(\theta/2)\sigma_\phi
\end{equation}
\begin{equation}
\sigma_\phi=\cos\phi\,\sigma_x+\sin\phi\,\sigma_y
\end{equation}
where $\theta$ is the rotation angle and $\phi$, the pulse phase, fixes the rotation axis in the $xy$-plane (note that $\phi$ is only defined up to multiples of $2\pi$).  As these are propagators a sequence of pulses must be written with time running from right to left, the reverse of the usual order for pulse sequences.  I will also use the notation $z_\alpha$ for a rotation by an angle $\alpha$ around the $z$-axis.

It is sometimes convenient to characterize such pulses by their propagator fidelity~\cite{Jones2011}
\begin{equation}
\mathcal{F}=\left|\textrm{tr}(U^\dag V)\right|/\textrm{tr}(U^\dag U)
\label{eq:fidelity}
\end{equation}
where $V$ is the propagator of the composite pulse in the presence of errors and $U$ is the desired unitary transformation; equivalently pulses can be categorized by their \textit{infidelity}, defined by $\mathcal{I}=1-\mathcal{F}$. (Note that this fidelity is essentially the Hilbert--Schmidt inner product between $U$ and $V$.)  One traditional approach is to expand the fidelity as a Taylor series in the size of the underlying error term, and then to seek to set as many low order coefficients to zero as possible \cite{Husain2013}. Alternatively one may isolate the error term in the propagator itself and then expand this as a Taylor series \cite{Brown2004,Brown2005,Alway2007}.

Starting from two identities (here and elsewhere I neglect physically meaningless global phases)
\begin{equation}
\pi_\beta\,\pi_\alpha=z_{2(\beta-\alpha)}\qquad z_\beta\,\pi_\alpha=\pi_\alpha\,z_{-\beta}=\pi_{\alpha+\beta/2}
\end{equation}
it can be immediately deduced that any sequence of $n$ individual $\pi$ rotations corresponds either to some $z$-rotation (for even $n$) or  to some $\pi_\phi$ rotation (for odd $n$).  In the latter case
\begin{equation}
\pi_{\phi_n}\,\pi_{\phi_{n-1}}\dots\pi_{\phi_1}=\pi_\Phi\qquad\Phi=\sum_{j=1}^{n}(-1)^{(j+1)}\phi_j
\end{equation}
and so to implement a \NOT\ gate the individual pulse phases should be chosen such that $\Phi=0$.

\section{First order errors}
As a simple example consider \textit{pulse strength errors}, which occur when the strength of the driving field used to induce transitions between qubit states deviates from its nominal value by some fraction $\epsilon$. In this case the rotation angle $\theta$ of each pulse is also increased by some fraction $\epsilon$ so that
\begin{equation}
V=[(1+\epsilon)\pi]_\phi=\pi_\phi\,(\epsilon\pi)_\phi
\end{equation}
giving
\begin{equation}
U^\dag V=(\pi\epsilon)_\phi=\openone-\epsilon(\I\pi/2)\sigma_\phi+O(\epsilon^2)
\end{equation}
so that the single pulse contains a first order error term.  The corresponding fidelity is given by
\begin{equation}
\mathcal{F}=|\cos(\epsilon\pi/2)|=1-\epsilon^2\pi^2/8+O(\epsilon^4)
\end{equation}
and so the pulse has a second order infidelity term.  In general a pulse with an error term of order $n$ will have infidelity of order $2n$.

In the presence of pulse strength errors the total propagator can be written as
\begin{equation}
V=\pi_{\phi_n}\,\delta_{\phi_n}\,\pi_{\phi_{n-1}}\delta_{\phi_{n-1}}\,\dots\pi_{\phi_1}\delta_{\phi_1}
\end{equation}
with $\delta=\epsilon\pi$.  Any composite pulse of this form will equal the desired propagator up to order zero as long as the phases are chosen such that $\Phi=0$, but to consider the first order error term it is more convenient to isolate the error term from the ideal evolution.  The identity
\begin{equation}
\theta_\beta\,\pi_\alpha=\pi_\alpha\,\theta_{2\alpha-\beta}
\label{eq:pimove}
\end{equation}
allows the $\pi$ pulses to propagated to one end of the sequence
\begin{equation}
V=\pi_{\phi_n}\,\pi_{\phi_{n-1}}\dots\pi_{\phi_1}\,\delta_{\phi'_n}\,\delta_{\phi'_{n-1}}\dots\delta_{\phi'_1}
\end{equation}
where the modified phases $\phi'_j$, known in NMR as the phases in the interaction frame \cite{Levitt1986} or toggling frame \cite{Suter1987a,Odedra2012b}, are given by
\begin{equation}
\phi'_j=(-1)^{(j+1)}\phi_j+\sum_{k<j}(-1)^{(k+1)}2\phi_k.
\end{equation}
Alternatively these phases can be described recursively
\begin{equation}
\phi'_{j+1}-\phi'_{j}=(-1)^{j}(\phi_{j+1}-\phi_j)\qquad\phi'_1=\phi_1
\end{equation}
to give a form which is more useful in some cases.

\begin{figure}
\includegraphics{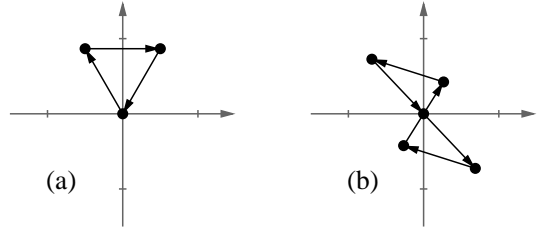}
\caption{Vector diagrams showing the cancellation of first order pulse strength error terms in composite pulse sequences with (a) $n=3$ and (b) $n=5$.  The equilateral polygons can be traced in either direction, depending on the precise relationship chosen between phases in the toggling frame, but the overall orientation is fixed by the constraint on $\Phi$.}\label{fig:polygons}
\end{figure}

Expanding the error terms to first order
\begin{equation}
\delta_\alpha\approx\openone-\I(\delta/2)\sigma_\alpha+O(\delta^2)
\end{equation}
allows the combined error term to be approximated as
\begin{equation}
\delta_{\phi'_n}\,\delta_{\phi'_{n-1}}\dots\delta_{\phi'_1}=\openone-\I(\delta/2)\sum_j\sigma_{\phi'_j}+O(\delta^2)
\end{equation}
and so a composite pulse which corrects first order pulse strength errors can be found by choosing the $\phi'_j$ such that
\begin{equation}
\sum_j\sigma_{\phi'_j}=0.
\label{eq:sumphi1}
\end{equation}
Since the operators in this sum all lie in the $xy$-plane and all have the same size, they can be mapped onto two-dimensional unit vectors and the sum solved geometrically: the underlying vectors must form a closed equilateral polygon of order $n$, as shown in Fig.~\ref{fig:polygons}.

\begin{figure*}
\includegraphics{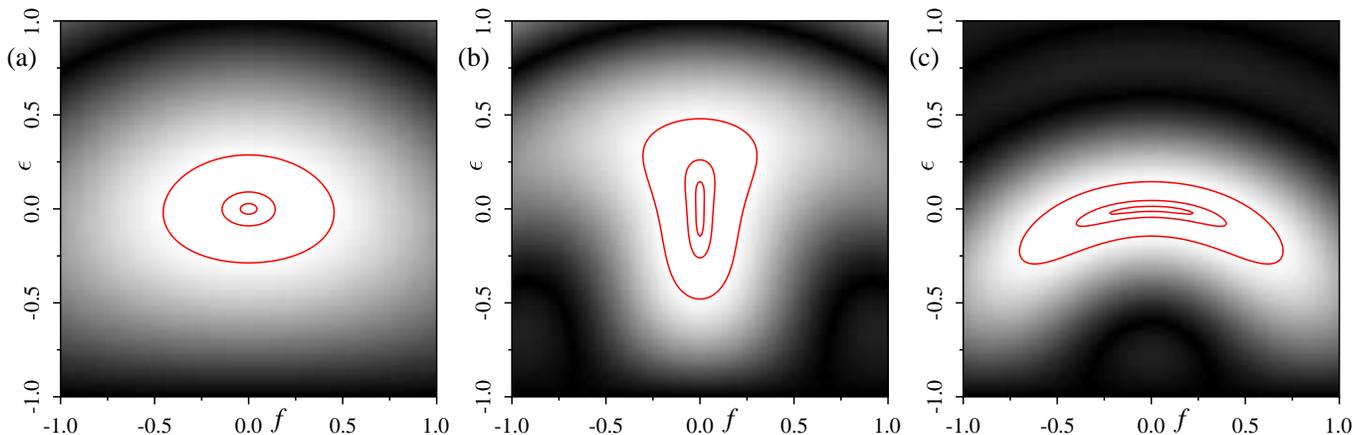}
\caption{(Color online) Fidelity achieved by (a) a simple pulse and composite pulses with $n=3$ optimized to suppress first-order pulse strength errors (b) and off-resonance errors (c).  Fidelity is plotted as a function of the fractional pulse strength error $\epsilon$ and the off-resonance fraction $f$.  Contours are drawn at 90\%, 99\% and 99.9\% fidelity, that is logarithmically spaced infidelities with the inmost contour at an infidelity of $10^{-3}$.}\label{fig:fNOB}
\end{figure*}

The geometric approach is particularly effective when $n=3$ as the equilateral triangle is uniquely defined up to rotations and reflections.  In particular such triangles require that
\begin{equation}
\phi'_3=\phi'_2\pm2\pi/3\quad\textrm{and}\quad\phi'_2=\phi'_1\pm2\pi/3
\end{equation}
where the two $\pm$ signs must be chosen with the same sense.  Together with the requirement that $\Phi=0$ these lead to three simultaneous equations which can be written as
\begin{equation}
\begin{pmatrix}0&-1&1\\1&-1&0\\1&-1&1\end{pmatrix}\begin{pmatrix}\phi_1\\\phi_2\\\phi_3\end{pmatrix}=\begin{pmatrix}\pm2\pi/3\\\pm2\pi/3\\0\end{pmatrix}
\end{equation}
with solutions
\begin{equation}
\mbox{\boldmath$(\phi)$}=(\phi_1,\, \phi_2,\, \phi_3)=(\mp2\pi/3,\,\mp4\pi/3,\,\mp2\pi/3).
\end{equation}
A second pair of solutions can be found by using an alternative third constraint, namely $\Phi=\pi$; this relies on the observation that $\pi_\pi$=$\pi_0$ up to irrelevant global phases.  All four of these pulse sequences have the same fidelity
\begin{equation}
\mathcal{F}=
1-3\pi^4\epsilon^4/128+O(\epsilon^6)
\end{equation}
and are entirely equivalent to each other.

The second important type of systematic error in NMR pulses is off-resonance errors.  These occur when the driving field is not exactly in resonance with the underlying transition between the spin states.  In this case
\begin{equation}
V(\theta,\phi)=\exp[-\I\theta(\sigma_\phi+f\sigma_z)/2]
\end{equation}
where $\theta$ and $\phi$ now describe the behaviour of the pulse \textit{on-resonance} and $f$, the off-resonance fraction, is the ratio between the frequency offset and the rotation frequency induced by the driving field.  In this case the situation is more complex than was seen for pulse strength errors, as the propagator is no longer easily divided into desired and undesired terms.  However it is possible to expand the propagator to first order, giving for a $\pi$ pulse
\begin{equation}
V(\pi,\phi)\approx\pi_\phi\lambda_{\phi+\pi/2}
\label{eq:por1}
\end{equation}
with $\lambda=2f$.  As before, the $\pi$ pulses can be propagated to one end of a composite pulse sequence
\begin{equation}
V=\pi_{\phi_n}\,\pi_{\phi_{n-1}}\dots\pi_{\phi_1}\,\lambda_{\phi''_n}\,\lambda_{\phi''_{n-1}}\dots\lambda_{\phi''_1}
\end{equation}
where the modified phases are now given by
\begin{equation}
\phi''_j=\phi'_j+(-1)^{(j+1)}\,\pi/2.
\end{equation}
As before a composite pulse which corrects first order off-resonance errors can be found by choosing the $\phi''_j$ such that
\begin{equation}
\sum_j\sigma_{\phi''_j}=0
\end{equation}
with the underlying vectors forming an equilateral polygon.  Once again the case $n=3$ is easily solved, to give $\phi_1=\phi_3=\pi/3$ and $\phi_2=2\pi/3$, and three other closely related solutions, which have fidelity
\begin{equation}
\mathcal{F}=1-\left(\frac{3+\pi^2}{8}\right)f^4+O(f^6).
\end{equation}

These composite pulses have been known for many years: see for example \cite{Tycko1985,Odedra2012b}, and note that the sequence for correcting pulse strength errors was subsequently generalized as the SCROFULOUS family of composite pulses \cite{Cummins2003}.  The properties of the first order error correcting sequences derived so far are summarized in Fig.~\ref{fig:fNOB}.  For applications in conventional NMR it is usually desirable to have moderately effective error suppression at large errors, but for applications in QIP it is necessary to achieve very precise rotations for small or moderate errors.  For this reason, one should concentrate on the area enclosed by the inmost contour line, corresponding to 99.9\% fidelity.  Note that in conventional NMR treatments the first-order error discussed here is referred to as the zero-order error, and this numbering convention continues at higher orders.

Examining the inner contour lines it is immediately clear that it is possible to greatly broaden the high precision region in one dimension, but that this is bought at some cost in increased sensitivity in the other dimension.  Later I will consider the possibility of either creating even greater precision in one dimension or of broadening this region in \textit{both} dimensions.  Both of these aims will require sequences with at least five pulses.

\section{Geometry}
It is instructive to consider an alternative ``brute-force'' approach for tackling this problem.  Consider a general sequence of three pulses with $\Phi=0$
\begin{equation}
\mbox{\boldmath$(\phi)$}=(\alpha,\,\beta,\,\beta-\alpha).
\end{equation}
In the presence of pulse strength errors the fidelity of such a composite pulse has the form
\begin{equation}
\mathcal{F}=1-\frac{\pi^2\epsilon^2}{8}\left[3+2\cos\alpha+2\cos(\alpha-\beta)+2\cos\beta\right]+O(\epsilon^4)
\end{equation}
and a good composite pulse can be found by minimising the trigonometric term (note that the form of the fidelity expression guarantees that this term will be non-negative, and so seeking to minimize it is equivalent to seeking to set it to zero).  Plotting it as a function of $\alpha$ and $\beta$ indicates that the minima lie along the line $\beta=-\alpha$, and imposing this constraint allows the solution to be easily located at $\alpha=\pm2\pi/3$.

Although this might seem quite different from the previous approach, they are in fact very similar.  The second order infidelity term in the Taylor series is simply given by half the magnitude of the first order error term in the propagator.  Clearly setting this term to zero and setting its magnitude to zero achieves the same effect.  However looking directly at the individual error terms allows geometric insights to be used, greatly reducing the complexity of the calculations.  This advantage becomes less clear once higher order terms are considered, and a judicious combination of geometry and algebra may be the best way to proceed.

The role of geometry in error tolerance was recently discussed by Ichikawa \textit{et al.} \cite{Ichikawa2012}. They were interested in distinguishing between Class A composite pulses and point-to-point pulses by interpreting Class A pulses as geometric quantum gates, that is gates where the dynamic phase vanishes for every eigenstate of the ideal gate operator \cite{Ota2009,Ota2009a,Kondo2011,Ichikawa2011,Ichikawa2012}.  They showed \cite{Ichikawa2012} that for composite pulses where the first order error defined above vanishes the dynamic phase also vanishes, and so every Class A composite pulse must also be a geometric quantum gate.

This observation only applies, however, to the suppression of the first order error term, and it is necessary to consider separately how higher order errors can be suppressed.  Note also that even for Class A pulses the error suppression will not be the same for every initial state: the composite pulse described above suppresses the first order pulse strength error for every initial state, but also suppresses the second order error for basis states.  In other words a composite pulse can be better as an inversion pulse than as a general rotor.  The fidelity measure, Eq.~\ref{eq:fidelity}, in effect reports the fidelity for the worst case initial states.

The geometric approach was also adopted by Merrill and Brown \cite{Merrill2012}, who noted the geometrical vector interpretation of the first order error terms.  They then proceed, however, to interpret higher order terms geometrically by considering the underlying Lie algebra.

\section{Symmetry}
The composite pulse sequences derived above are \textit{time symmetric}, that is the sequence of pulse phases is the same when reversed.  The fidelity plots in Fig.~\ref{fig:fNOB} are also symmetric around the line $f=0$.  These two facts are related: it can be shown that for any time symmetric sequence of $\pi$ pulses the fidelity is an even function of the off-resonance fraction $f$, containing only even powers in its Taylor expansion.

In general, however, composite pulses need not be symmetric, and in this case the fidelity function will contain odd powers in its Taylor expansion, and the fidelity plot will not be symmetric.  This asymmetric plot can be reversed around $f=0$ either by reversing the order of all the pulses, or by negating the phases of all the pulses.  For a \textit{time antisymmetric} composite pulse, with $\phi_n=-\phi_1$ and so on, these two operations are equivalent.  For a time symmetric pulse they are different, but neither affects the symmetric fidelity function \cite{Levitt2008}.

The response to pulse strength errors is different.  In the absence of off-resonance errors the fidelity of a sequence of $\pi$ pulses is always an even function of the fractional pulse strength error $\epsilon$.  This can be seen by noting that a negative value of $\epsilon$ is equivalent to a positive value of $\epsilon$ combined with a shift in the pulse phase by $\pi$.  In the presence of off-resonance errors, however, this argument breaks down, and no particular symmetry is found.

The symmetry of composite pulses also has effects in the toggling frame: a composite pulse with antisymmetric phases $\phi_j$ always has symmetric phases $\phi'_j$, a fact that will be of some importance later.  The converse does not quite apply:  a composite pulse with symmetric phases $\phi_j$ need not have antisymmetric phases $\phi'_j$ in general, but if the pulse has $\Phi=0$, so that it is correct to zero order, then it \textit{will} do so.  As I only consider pulses which are correct to zero order this relationship will be assumed from now on.  Note that antisymmetric sequences of $\pi$ pulses \textit{always} have $\Phi=0$, and so provide plausible candidates for composite pulse \NOT\ gates \cite{Husain2013}.

For an antisymmetric pulse the central phase is fixed at $\phi_{(n+1)/2}=0$, while for a symmetric pulse it is apparently free; however the requirement that $\Phi=0$ fixes this value, and so both symmetric and antisymmetric composite pulses have $(n-1)/2$ free phases which can be adjusted.  The more general non-symmetric composite pulses have $n-1$ free phases, twice as many.  However the simpler symmetric and antisymmetric pulses have many advantages and will be seen frequently in the sections that follow.  Antisymmetric composite pulses have particular advantages in some conventional NMR experiments \cite{Odedra2012,Odedra2012a}, where it is possible to render effectively invisible any signal arising from certain types of error, so that these errors lead to a drop in overall signal strength but do not generate any erroneous signal terms.  This phenomenon is not normally useful, however, in QIP experiments, where symmetric composite pulses are often preferred.

\section{Second order corrections}
The composite pulses described above provide the best correction of pulse strength errors possible with a sequence of three pulses, but it is possible to also remove second order errors with a sequence of five pulses.  The most obvious approach is to expand the error propagator up to second order, but this is not ideal as the second order terms in this non-linear propagator include contributions from the first order error \cite{Odedra2012b}, and it is convenient to separate these from the genuinely second order terms.

\begin{figure*}
\includegraphics{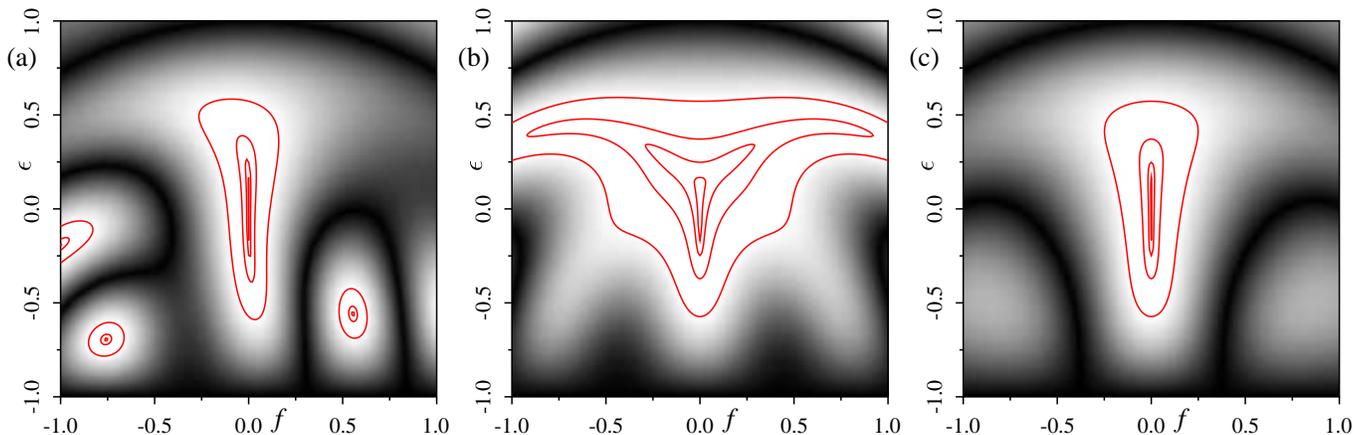}
\caption{(Color online) Fidelity achieved by some composite pulses designed to suppress pulse strength errors: (a) \Fn1, (b) symmetric BB1, (c) a new symmetric pulse with $n=5$.  The inmost contour is drawn at an infidelity of $10^{-4}$. }\label{fig:FBS}
\end{figure*}

The traditional approach in the NMR community is to use average Hamiltonian theory \cite{Haeberlen1968,Tycko1984,Tycko1985a,EBWbook,Odedra2012b}, but for composite pulses it can be simpler to proceed directly from the underlying Baker--Campbell--Hausdorff (BCH) relation \cite{EBWbook,Campbell1897}
\begin{equation}
\e^B\e^A=\exp\{B+A+{\textstyle\frac{1}{2}}[B,A]
+\dots\}
\end{equation}
where terms above second order have been dropped.  This can be extended in the obvious way, and for pulse strength errors leads to
\begin{equation}
\delta_{\phi'_n}\,\delta_{\phi'_{n-1}}\dots\delta_{\phi'_1}=\exp\{\Delta_1+\Delta_2+\dots\}
\end{equation}
with
\begin{equation}
\Delta_1=-\I\frac{\delta}{2}\sum_j\sigma_{\phi'_j}
\label{eq:pse1}
\end{equation}
and
\begin{align}
\Delta_2&=-\frac{\delta^2}{8}\sum_j\sum_{k<j}\big[\sigma_{\phi'_j},\sigma_{\phi'_k}]\\
&=\I\frac{\delta^2}{4}\sum_j\sum_{k<j}\sin(\phi'_j-\phi'_k)\,\sigma_z.
\end{align}
We have already seen how to set $\Delta_1=0$, removing the first order error, while to remove the second order error requires that
\begin{equation}
\sum_j\sum_{k<j}\sin(\phi'_j-\phi'_k)=0.
\label{eq:pse2}
\end{equation}

A traditional NMR approach to achieve this is to choose an antisymmetric pulse sequence, with a symmetric sequence of phases in the toggling frame, as this symmetry forces the terms in $\Delta_2$ to cancel.  Thus \textit{any antisymmetric composite pulse which suppresses the first order pulse strength error will automatically suppress the second order error}.  It remains to be shown that an antisymmetric composite pulse with no first order error can be found in the case $n=5$.  In fact a unique pair of solutions exist, with $\phi_1=3\psi$, $\phi_2=\psi$ and $\psi=\pm\arccos(-1/4)$, the well known \Fn1 pulse \cite{Wimperis1991,Husain2013} (note that the order of labelling phases 1 and 2 is reversed in some descriptions).  The vectors describing the corresponding toggling frame phases, $\phi'_1=\phi'_5=3\psi$, $\phi'_2=\phi'_4=5\psi$, and $\phi'_3=4\psi$, form an equilateral pentagon, but it is certainly not a regular pentagon.  In fact it is a stellated pentagon, with one side passing through a vertex, as shown in Fig.~\ref{fig:polygons}.

The \Fn1 composite pulse has a fidelity
\begin{equation}
\mathcal{F}=1-5\pi^6\epsilon^6/1024+O(\epsilon^8).
\end{equation}
This fidelity can be equalled by a number of other composite pulses, but no five pulse sequence is significantly more effective for suppressing pulse strength errors.  Its response in the presence of simultaneous pulse strength and off-resonance errors is shown in Fig.~\ref{fig:FBS}(a).  This plot shows the solution with a positive value of $\psi$; the second solution is simply the time-reversed (negative phase) equivalent, and has an equivalent (but mirror image) fidelity.  Note that in this figure the inmost contour is at 99.99\% fidelity, ten times better than the inmost contour in Fig.~\ref{fig:fNOB}.

\section{Reordering pulses}
The phase angles in the \Fn1 pulse are closely related to the better known BB1 pulse \cite{Wimperis1994,Cummins2003}, and here I consider the possibility of reordering sequences of $\pi_\phi$ pulses.  The fidelity definition, Eq.~\ref{eq:fidelity}, is invariant under cyclic reorderings of its components, and Eq.~\ref{eq:pimove} allows the ideal $\pi_0$ pulse to be moved through the fidelity definition as long as the corresponding phases are negated \cite{Husain2013}.

Taken together these results lead to the conclusion that in the presence of pulse strength errors the fidelity of the \Fn1 composite pulse
\begin{equation}
\mbox{\boldmath$(\phi_{F_1})$}=(3\psi,\,\psi,\,0,\,-\psi,\,-3\psi)
\end{equation}
is \textit{identical} to the fidelity of the reordered pulse sequence
\begin{equation}
\mbox{\boldmath$(\phi_{BB1})$}=(0,\,\psi,\,3\psi,\,3\psi,\,\psi).
\end{equation}
This argument does not, however, apply to off-resonance errors, and the response of the two pulse sequences above is very different.  Neglecting the initial $\pi_0$ pulse the BB1 pulse sequence comprises a nested set of $2\pi$ rotations, and to first order such rotations have no off-resonance error.  This can be seen either by expanding the single pulse propagator directly \cite{Husain2013}, or by using the first order expansion of a $\pi$ pulse, Eq.~\ref{eq:por1}, and noting that the second $\pi$ pulse generates a spin echo \cite{Hahn1950}, which removes the first order error terms.  As a consequence the first order response of a BB1 pulse to off-resonance errors arises solely from the initial $\pi_0$ pulse, and so the response is very similar to a simple pulse.  The similarity can be increased still further by splitting the $\pi_0$ pulse into two and moving one half to the end of the sequence \cite{Husain2013}.

This fully time symmetric BB1 sequence, with the propagator
\begin{equation}
(\pi/2)_0\,\pi_\psi\,(2\pi)_{3\psi}\,\pi_\psi\,(\pi/2)_0
\end{equation}
has a frequency symmetric response to off-resonance errors, so the fidelity is purely an even function of $f$.  The response of BB1 to simultaneous errors is compared with \Fn1 in Fig.~\ref{fig:FBS}(b).  Although BB1 does not actively suppress off-resonance errors, the increased sensitivity to such errors exhibited by \Fn1 is overcome by symmetrisation.

\section{Symmetric sequences}
While this BB1 composite pulse is time symmetric, it breaks the informal rule of considering only sequences of $\pi$ pulses.  It is interesting to consider whether an intrinsically time symmetric sequence of five $\pi$ pulses exists which suppresses pulse strength errors to second order.  In this case it is necessary to consider both the first order and second order errors, but these can be taken one at a time.  In the discussions above Eq.~\ref{eq:sumphi1} has been interpreted geometrically, but it can instead be considered as the pair of equations
\begin{equation}
\sum\cos(\phi'_j)=\sum\sin(\phi'_j)=0.
\end{equation}
For a time symmetric sequence the toggling frame phases will be antisymmetric, and so the sum of sin terms above will automatically be equal to zero.  Thus for the general symmetric sequence
\begin{equation}
\mbox{\boldmath$(\phi)$}=(\alpha,\,\beta,\,2\beta-2\alpha,\,\beta,\,\alpha).
\end{equation}
the first order error will be suppressed if
\begin{equation}
1+2\cos(\alpha)+2\cos(2\alpha-\beta)=0.
\end{equation}
This has the solution
\begin{equation}
\beta=2\alpha\pm\arccos[-(1+2\cos\alpha)/2]
\end{equation}
which is only defined when $\pi/3\le\alpha\le5\pi/3$ or $\alpha$ is offset from this range by multiples of $2\pi$.  The final phase $\alpha$ can then be varied in an attempt to minimize the second order error term, and it turns out that this term can in fact be entirely removed by choosing
\begin{equation}
\alpha=\mp2\arcsin(\sqrt[4]{5/32}).
\end{equation}
Note that the two $\pm$ signs must be chosen with the opposite signs as indicated above but the two solutions are fundamentally equivalent.  The fidelity of this composite pulse in the absence of off-resonance errors is \textit{identical} to that of \Fn1 and BB1.  However, as shown in Fig.~\ref{fig:FBS}(c), the fidelity in the presence of off-resonance errors, although symmetric, is similar to that of \Fn1 and much worse than that of BB1.  Explicit pulse phases are listed in Table~\ref{tab:phi5}.

\begin{table}
\begin{tabular}{lrrrrr}
\hline
&$\phi_1$&$\phi_2$&$\phi_3$&$\phi_4$&$\phi_5$\\
\hline
\ref{fig:FBS}(a)&\tdeg{313.4}&\tdeg{104.5}&\tdeg{0.0}&\tdeg{255.5}&\tdeg{46.6}\\
\ref{fig:FBS}(c)&\tdeg{77.9}&\tdeg{20.6}&\tdeg{245.4}&\tdeg{20.6}&\tdeg{77.9}\\
\ref{fig:FKBO}(a)&\tdeg{240.0}&\tdeg{210.0}&\tdeg{300.0}&\tdeg{210.0}&\tdeg{240.0}\\
\ref{fig:FKBO}(b)&\tdeg{8.7}&\tdeg{94.3}&\tdeg{300.0}&\tdeg{325.7}&\tdeg{111.3}\\
\ref{fig:FKBO}(c)&\tdeg{111.3}&\tdeg{145.7}&\tdeg{300.0}&\tdeg{274.3}&\tdeg{8.7}\\
\hline
\end{tabular}
\caption{Explicit pulse phases for composite pulses comprising five $\pi$ pulses and depicted in Figs. \ref{fig:FBS} and \ref{fig:FKBO}.}\label{tab:phi5}
\end{table}

\section{Off-resonance errors}
It is tempting to try to design an antisymmetric composite pulse which suppresses the second order off-resonance term in the same way as can be achieved for pulse strength errors, but this approach is not successful.  The pulse phases needed to suppress the first order off-resonance error are $\phi_1=\arccos(11/16)\approx46.6^\circ$ and $\phi_2=\arccos(1/4)\approx75.5^\circ$, with $\phi_3=0$, $\phi_4=-\phi_2$ and $\phi_5=-\phi_1$ as usual.  This pulse was briefly discussed by Odedra \textit{et al.} who noted its poor performance \cite{Odedra2012b}.  Evaluating the propagator for this pulse shows that it contains a second order error term, and (equivalently) the fidelity expression
\begin{equation}
{\mathcal{F}}=1-\pi^2f^4/2+O(f^5)
\end{equation}
is only correct to fourth order.

The reason for this behaviour is not hard to find: the second order error term for off-resonance errors is more complex than for pulse strength errors.  Expanding the single pulse propagator to second order gives
\begin{equation}
V(\pi,\phi)\approx\pi_\phi\lambda_{\phi+\pi/2}\mu_\phi
\label{eq:por2}
\end{equation}
with $\mu=\pi f^2/2$.  Applying the BCH relation in this case is fairly simple as the cross terms between these pulses are $O(f^3)$, and so can be neglected, leaving the forms
\begin{equation}
\Delta_1=-\I\frac{\lambda}{2}\sum_j\sigma_{\phi''_j}
\label{eq:ore1}
\end{equation}
as before, but
\begin{equation}
\Delta_2=\I\frac{\lambda^2}{4}\sum_j\sum_{k<j}\sin(\phi''_j-\phi''_k)\,\sigma_z-\I\frac{\mu}{2}\sum_j\sigma_{\phi'_j}
\end{equation}
where the left hand part of the expression is analogous to that for pulse strength errors, but the additional terms on the right hand side arise from the second order errors in the individual pulses.  Designing a composite pulse with second-order tolerance of off-resonance errors requires suppression of all these terms, which will not be achieved by an antisymmetric composite pulse: the additional term depends on $\phi'$, not $\phi''$, and so is not suppressed along with the first order term.

It is interesting to note, however, that the additional second order term takes the same form as the first order term for a pulse strength error.  Thus any composite pulse with second order tolerance of off-resonance errors must have first order tolerance of pulse strength errors, that is the composite pulse must have \textit{simultaneous tolerance} of both sorts of error.

\section{Simultaneous error tolerance}
To achieve simultaneous tolerance of pulse strength and off-resonance errors it is necessary to choose the composite pulse phases such that Eqns. \ref{eq:pse1} and \ref{eq:ore1} are both equal to zero.  This might seem difficult to achieve, but is in fact straightforward as noted by Tycko \textit{et al.} \cite{Tycko1984,Tycko1985a}.  The two sets of toggling frame phases, $\phi'_j$ and $\phi''_j$ are alternately different by $+\pi/2$ (for odd $j$) and $-\pi/2$ (for even $j$).  Thus if the phases are chosen such that Eq.~\ref{eq:pse1} separately sums to zero for the odd terms and the even terms, then Eq.~\ref{eq:ore1} will also sum to zero \cite{Odedra2012b}.  A little thought shows that this separate cancellation is \textit{necessary} as well as \textit{sufficient} for simultaneous error tolerance.

This can, in fact, be achieved with a sequence of five pulses.  The phase vectors describing the three odd-numbered pulses must form an equilateral triangle, while the vectors for the two even pulses must be antiparallel.  Equivalently the three odd phases, $\phi'_1$, $\phi'_3$ and $\phi'_5$, must differ by $\pm2\pi/3$, providing two constraints on the pulse phases, while the even phases, $\phi'_2$ and $\phi'_4$, must differ by $\pi$, providing an additional constraint.  Finally the requirement $\Phi=0$ provides a fourth constraint, and so one of the five phases, say $\phi_2$, can be varied at will.  Solving the simultaneous equations gives the family of solutions
\begin{equation}
\mbox{\boldmath$(\phi)$}=(\pi+2\alpha,\,\alpha,\,-\pi/3,\,-5\pi/3-\alpha,\,-7\pi/3-2\alpha)
\end{equation}
where $\alpha$ can be chosen at will.  This family is well known in the context of inversion pulses \cite{Tycko1984,Tycko1985a}, and has also been discussed as general rotors \cite{Odedra2012b}, but its properties have not been fully explored.  All members of the family suppress first order pulse strength and off-resonance errors, but the effect on second order errors depends on the value of $\alpha$.

Studies to date have largely concentrated on values of $\alpha$ which result in particularly simple pulse sequences.  For example one can set $\phi_1=\phi_2$ by choosing $\alpha=-\pi$.  Offsetting all phase angles by $\pi$ (reflecting the equivalence of $\pi_0$ and $\pi_\pi$ pulses), and rewriting everything in the range between $0$ and $2\pi$ gives
\begin{equation}
\mbox{\boldmath$(\phi)$}=(0,\,0,\,2\pi/3,\,\pi/3,\,2\pi/3)
\end{equation}
which is the $\textrm{S}_1$ inversion pulse \cite{Tycko1984,Tycko1985a}.  Alternatively one can set $\phi_1=\phi_5$ by choosing $\alpha=-5\pi/6$ to get the time symmetric sequence
\begin{equation}
\mbox{\boldmath$(\phi)$}=(4\pi/3,\,7\pi/6,\,5\pi/3,\,7\pi/6,\,4\pi/3).
\end{equation}
Although this composite pulse may not appear familiar, it is in fact closely related to the Knill pulse \cite{Ryan2010} which is frequently used for dynamic decoupling \cite{Souza2011b,Souza2012}.  While the original Knill pulse is frequently described as a $\pi_0$ pulse followed by a $z$-rotation, it is better thought of as $\pi$ rotation around some other axis in the $xy$-plane.

\begin{figure}
\includegraphics{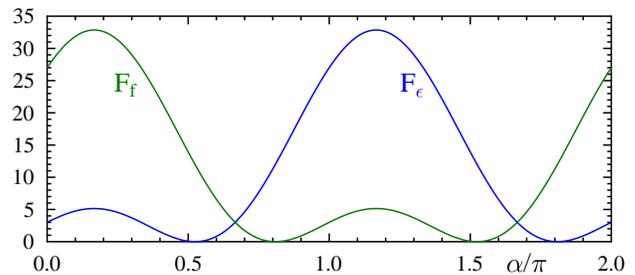}
\caption{(Color online) Coefficient of the fourth order infidelity terms for pulse strength error ($F_\epsilon$) and off-resonance error ($F_f$) in five pulse composite pulses with simultaneous compensation of both errors to first order.}\label{fig:FeFf}
\end{figure}
\begin{figure*}
\includegraphics{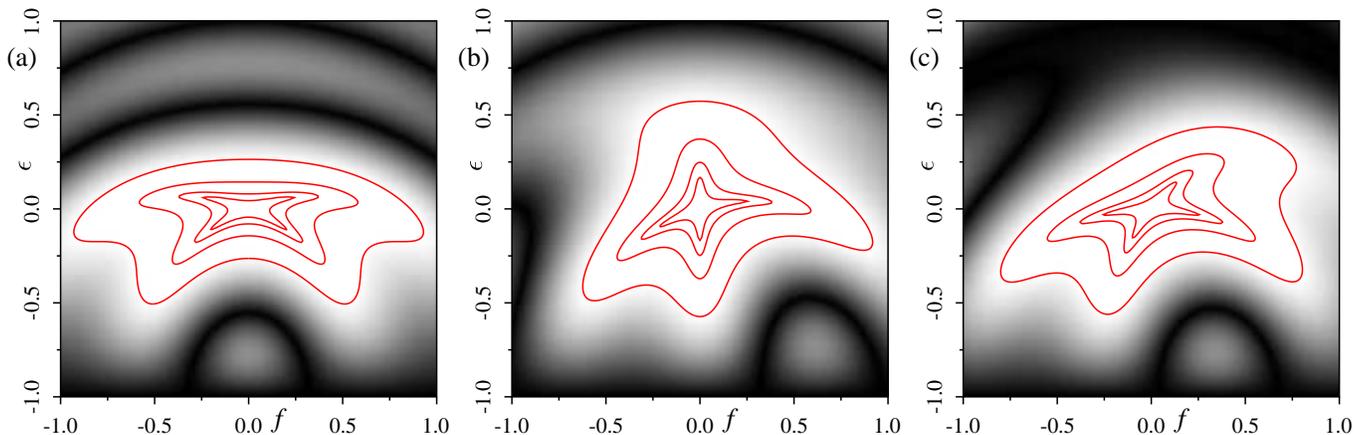}
\caption{(Color online) Fidelity achieved by some composite pulses with $n=5$ designed to suppress both pulse strength and off-resonance errors: (a) the symmetric Knill type pulse, (b) pulse with suppression of second order pulse strength errors, (c) pulse with suppression of second order off-resonance errors.  The inmost contour is drawn at an infidelity of $10^{-4}$. }\label{fig:FKBO}
\end{figure*}

A more interesting approach is to find the propagator fidelity of the family of pulses under both pulse strength and off-resonance errors.  Expanding these to fourth order gives
\begin{equation}
{\mathcal F_\epsilon}=1-\frac{1}{8}\left(\frac{\pi\epsilon}{2}\right)^4F_\epsilon+O(\epsilon^6)
\label{eq:Fe}
\end{equation}
and
\begin{equation}
{\mathcal F_f}=1-\frac{1}{8}f^4F_f+O(f^5)
\label{eq:Ff}
\end{equation}
where the coefficients of the fourth order terms are
\begin{equation}
F_{\epsilon,f}=11\mp12\cos\alpha+4\cos2\alpha+4\sqrt3(\sin2\alpha\mp\sin\alpha)
\label{eq:F5ef}
\end{equation}
and the $\mp$ symbols are minus signs for $F_\epsilon$ and plus signs for $F_f$.  In fact these are essentially the same function, with $\alpha$ simply offset by $\pi$.  (Note that the different pre-factors in Eqns \ref{eq:Fe} and \ref{eq:Ff} reflects the different way in which the two errors are parameterized.)  As the graphs in Fig.~\ref{fig:FeFf} show it is possible to have very significant control over the size of these error coefficients.  Interestingly the two pulses discussed above appear to use quite poor choices of $\alpha$; in particular the Knill pulse occurs at a global maximum in the pulse strength error and a local maximum in the off-resonance error.  However the Knill pulse does have the advantage of being time symmetric, and so its response to off-resonance errors is frequency symmetric, as shown in Fig.~\ref{fig:FKBO}.

There are obvious alternative approaches.  First, consider the points where the two curves cross, which occur at $\alpha=2\pi/3$ and $\alpha=5\pi/3$.  At these points the sensitivities of the composite pulse to pulse strength and off-resonance errors are in some sense balanced.  More interestingly, note that the two curves touch the $x$-axis, and so it is possible to find values of $\alpha$ such that one of the two coefficients is set to zero.  This corresponds to finding a pulse sequence which eliminates the second order term in the corresponding error.  The solution for removing second order pulse strength errors occurs at
\begin{equation}
\alpha=\pm\arccos[(3\mp\sqrt{13})/8]\approx94.3^\circ\quad\textrm{or}\quad325.7^\circ
\end{equation}
while for removing second order off-resonance error the solution occurs at
\begin{equation}
\alpha=\pm\arccos[(-3\mp\sqrt{13})/8]\approx145.7^\circ\quad\textrm{or}\quad274.3^\circ.
\end{equation}
The two choices in each case give equivalent composite pulses; note that as expected from the discussion below Eq.~\ref{eq:F5ef} sequences optimised for pulse strength errors and off-resonance errors have values of $\alpha$ differing by $180^\circ$.  Fidelity plots for these new composite pulses are shown in Fig.~\ref{fig:FKBO}, confirming the expected behaviour, while some explicit pulse phases are listed in Table~\ref{tab:phi5}.

These composite pulses also illustrate some general principles.  It is possible to remove first and second order pulse strength errors without using an antisymmetric pulse: antisymmetry is a sufficient but not a necessary condition.  This is important, as it is easy to show that no antisymmetric composite pulse with $n=5$ can suppress both pulse strength and off-resonance errors: antisymmetric pulses must have $\phi'_2=\phi'_4$, which contradicts the requirement that these toggling frame phases must be separated by $\pi$.  Once antisymmetry is abandoned it is possible to develop a composite pulse with $n=5$ which suppresses first and second order off-resonance errors, and this composite pulse \textit{also} suppresses first order pulse strength errors.  

Finally I return to the different parametrization of pulse strength and off-resonance errors briefly discussed above.  This is a general phenomenon, and as a result the inner contours in fidelity plots will be a factor of $\pi/2$ wider in composite pulses optimized for off-resonance errors than in corresponding pulses optimized for pulse strength errors.  Similarly the behaviour of the outer contours is dominated by the simple observation that, in the absence of off-resonance errors, the fidelity must always fall to zero at $\epsilon=\pm1$, while the equivalent limits for off-resonance errors occur at $f=\pm\sqrt{3}$.  For this reason fidelity plots for high order composite pulses will always look better in the off-resonance dimension than in the pulse strength dimension.

\section{Sequences with seven pulses}
\begin{figure*}
\includegraphics{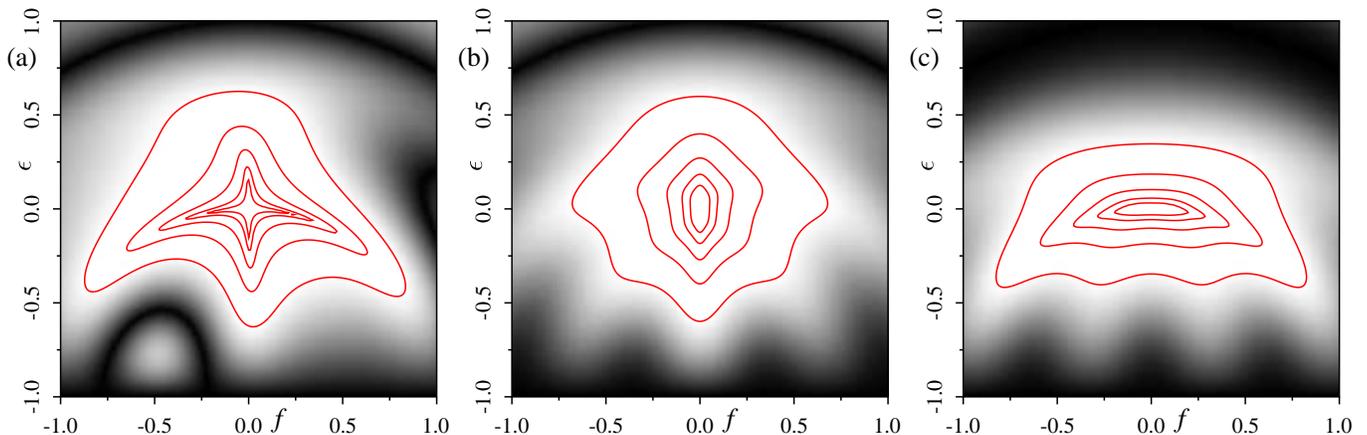}
\caption{(Color online) Fidelity achieved by various composite pulses with $n=7$: (a) second order suppression of both pulse strength and off-resonance errors, (b) symmetric pulse optimized for pulse strength errors, (c) symmetric pulse optimized for off-resonance errors.  The inmost contour is at an infidelity of $10^{-5}$. }\label{fig:fBBOO}
\end{figure*}

This approach can be extended to longer pulses, and remains fairly straightforward for the case $n=7$.  In this case there are three even pulses, which must form an equilateral triangle in the toggling frame, and four odd pulses which must form either a rhombus or a degenerate arrowhead.  The case of a rhombus leads to the two constraints $\phi'_5=\phi'_1+\pi$ and $\phi'_7=\phi'_3+\pi$, while the case of an arrowhead requires $\phi'_3=\phi'_1+\pi$ and $\phi'_7=\phi'_5+\pi$.  Together with the two constraints on the even pulses and the usual constraint that $\Phi=0$ this leaves two free parameters which can be adjusted to fine-tune the pulse sequence.

Solving for the case of the rhombus with two free parameters leads to the phases
\begin{multline}
\mbox{\boldmath$(\phi)$}
=\left(\alpha,\,2\beta,\,\beta,\,-2\pi/3,\,-\pi/3-\alpha+2\beta,\right.\\
\left.-2\pi/3-2\alpha+4\beta,\,-\pi-2\alpha+3\beta\right)
\end{multline}
where $\alpha$ and $\beta$ can be chosen at will.  As before the fidelity can be written as a Taylor series in $\epsilon$ and $f$, with fourth order coefficients that are functions of both $\alpha$ and $\beta$.  A pulse with simultaneous suppression of second order pulse strength and off-resonance errors can be found by solving for $F_\epsilon=F_f=0$, but as both coefficients are non-negative it is sufficient to solve the simpler equation $F_\epsilon+F_f=0$.  Even this function is difficult to solve analytically, but it can be easily investigated numerically.

Plotting $F_\epsilon+F_f$ as a function of $\alpha$ and $\beta$ reveals four distinct minima where the function is equal to zero.  These all occur at points where $\beta=\alpha-2\pi/3$, and imposing this constraint makes the equation easy to solve.  The solutions occur at
\begin{equation}
\alpha=\pm\arccos\left(\pm\frac{1}{2}\sqrt{\frac{4\pm\sqrt{13}}{2}}\right)
\end{equation}
where the first two $\pm$ signs must take the same value, but the third sign can be varied independently.

These four roots come in two pairs, which lead to equivalent pulse sequences, so there are only two genuinely distinct solutions.  Of these two solutions one is much better than the other, as it has smaller higher order errors.  The best result occurs at $\alpha\approx192.8^\circ$, and its behaviour is shown in Fig.~\ref{fig:fBBOO}(a); the explicit pulse phases are given in Table~\ref{tab:phi7}.  These solutions were obtained assuming that the even phases in the toggling frame \textit{increase} around the equilateral triangle, but the results are entirely equivalent if decreasing angles are used instead.

\begin{table}
\begin{tabular}{lrrrrrrr}
\hline
&$\phi_1$&$\phi_2$&$\phi_3$&$\phi_4$&$\phi_5$&$\phi_6$&$\phi_7$\\
\hline
\ref{fig:fBBOO}(a)&\tdeg{192.8}&\tdeg{145.7}&\tdeg{72.8}&\tdeg{240.0}&\tdeg{252.8}&\tdeg{145.7}&\tdeg{12.8}\\
\ref{fig:fBBOO}(b)&\tdeg{252.5}&\tdeg{265.0}&\tdeg{97.5}&\tdeg{170.0}&\tdeg{97.5}&\tdeg{265.0}&\tdeg{252.5}\\
\ref{fig:fBBOO}(c)&\tdeg{72.5}&\tdeg{265.0}&\tdeg{277.5}&\tdeg{170.0}&\tdeg{277.5}&\tdeg{265.0}&\tdeg{72.5}\\
\hline
\end{tabular}
\caption{Explicit pulse phases for composite pulses comprising seven $\pi$ pulses and depicted in Fig.~\ref{fig:fBBOO}.}\label{tab:phi7}
\end{table}

Next consider the case where the four odd pulses form an arrowhead in the toggling frame.  Once again the equations can be solved with two free parameters, but in this case it is not possible to achieve simultaneous suppression of the second order pulse strength and off-resonance errors, and I do not pursue this possibility further.

The behaviour of the symmetric Knill pulse in the case $n=5$ suggests that there is some value in imposing overall symmetry on the composite pulse sequence.  This can be achieved by imposing a single additional constraint, as symmetrising any pair of pulses is sufficient, together with the other constraints, to symmetrize the entire sequence.  Symmetric pulse sequences have antisymmetric phases in the toggling frame, and so $\phi'_5=-\phi'_3$.  This collapses the distinction between rhombus and arrowhead sequences, and the solutions can be found by requiring that $\phi'_3=\phi'_1\pm\pi$.  For the case of the plus sign the solution is
\begin{multline}
\mbox{\boldmath$(\phi)$}
=(\alpha,\,2\pi/3+2\alpha,\,7\pi/3+3\alpha,\,10\pi/3+4\alpha,\\
7\pi/3+3\alpha,\,2\pi/3+2\alpha,\,\alpha)
\end{multline}
and the free parameter $\alpha$ can be varied in order to optimize the suppression of either pulse strength or off-resonance errors.  In particular, choosing
\begin{equation}
\alpha=-\arccos\left[\left(3-\sqrt{61}\right)/16\right]
\end{equation}
completely removes the second order pulse strength error, while choosing
\begin{equation}
\alpha=\arccos\left[\left(\sqrt{61}-3\right)/16\right]
\end{equation}
completely removes the second order off-resonance error.  The wider behaviour of these pulses is shown in Fig.~\ref{fig:fBBOO}(b) and (c), and explicit pulse phases are given in Table~\ref{tab:phi7}.  As usual these plots show better visible performance for off-resonance errors than for pulse strength errors, and values of $\alpha$ for pulse sequences optimised for off-resonance errors differ from those optimised for pulse strength errors by $180^\circ$.

For the case of a minus sign, so that $\phi'_3=\phi'_1-\pi$, the general solution is
\begin{multline}
\mbox{\boldmath$(\phi)$}
=(\alpha,\,2\pi/3+2\alpha,\,\pi/3+\alpha,\,-2\pi/3,\\
\pi/3+\alpha,\,2\pi/3+2\alpha,\,\alpha)
\end{multline}
As before it is possible to find sequences optimized for either pulse strength or off-resonance error, but in this case optimising for one sort of error seems to lead to large second order error terms of the other kind (they do, of course, continue to suppress all first order errors).  The resulting composite pulses thus quite sharply favour one type of error over the other, and are not considered further here.

Finally I consider what can be achieved with antisymmetric seven pulse sequences.  As for $n=5$ it is not possible to simultaneously remove first order pulse strength and off-resonance errors, as the symmetry of the toggling frame phases prevents this.  In particular the requirement that $\phi'_6=\phi'_2$ prevents the formation of an equilateral triangle from the even phases.

\section{Sequences with nine pulses}
Tackling the general case with $n=9$ is significantly more difficult than the lower numbers, as it is now difficult to use geometric insight to make initial progress.  The four even pulses must form a rhombus or arrowhead in the toggling frame, but the five odd pulses must form a pentagon, for which there are no simple geometric restrictions.  It is, however, possible to make some progress by imposing particular symmetries on the problem.

The antisymmetric case was studied by Odedra \textit{et al.} \cite{Odedra2012b}.  An antisymmetric pulse with $n=9$ has only four controllable phases, and as the toggling frame phases must be symmetric there is no distinction between the rhombus and arrowhead cases.  A solution, which they call \mbox{ASBO-9}, occurs when
\begin{multline}
\mbox{\boldmath$(\phi)$}
=(4\alpha+\psi,\,3\alpha+2\psi,\,2\alpha+\psi,\,\alpha+\pi,\,0,\\
-\alpha-\pi,\,-2\alpha-\psi,\,-3\alpha-2\psi,\,-4\alpha-\psi)
\end{multline}
where $\psi=\arccos(-1/4)$ as usual, and $\alpha$ can be varied at will  \cite{Odedra2012b}.  All such composite pulses suppress both first and second order error terms arising from both pulse strength and off-resonance errors, and higher order error terms can be partially controlled by the choice of $\alpha$.

\begin{figure*}
\includegraphics{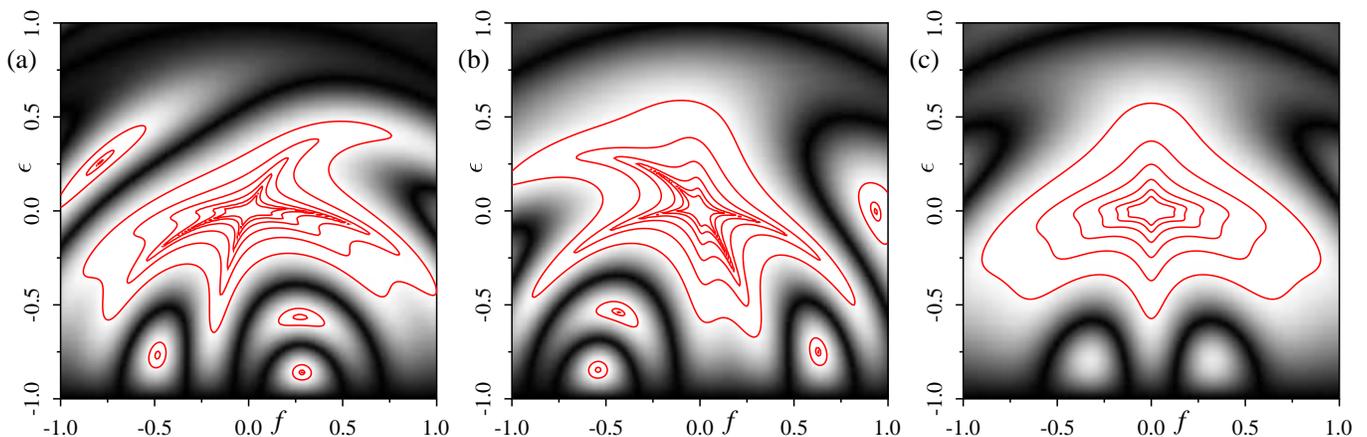}
\caption{(Color online) Fidelity achieved by various composite pulses with $n=9$ giving simultaneous suppression of first and second order pulse strength and off-resonance errors: (a) \mbox{ASBO-9(7A)}, (b) \mbox{ASBO-9(7B)}, (c) a new symmetric pulse.  The inmost contour is at an infidelity of $10^{-6}$. }\label{fig:fBO9}
\end{figure*}

It is not possible to completely remove either third order error term by varying $\alpha$, but the third order pulse strength error can be minimized by choosing $\alpha\approx308.0^\circ$, while the off-resonance term can be minimized by choosing $\alpha\approx128.0^\circ$; these values of $\alpha$ differ by $180^\circ$ as usual.  Alternatively the two error terms can be balanced by choosing $\alpha=\pm\psi$.  The first two solutions correspond to the pulse sequences \mbox{ASBO-9($\textrm{B}_1$)} and \mbox{ASBO-9($\Omega$)}, while the latter two are the pulse sequences \mbox{ASBO-9(7A)} and \mbox{ASBO-9(7B)} respectively \cite{Odedra2012b}.  The performance of the last two pulse sequences is shown in Fig.~\ref{fig:fBO9}(a) and (b), and the phase angles are listed in Table~\ref{tab:phi9}.

\begin{table}
\begin{tabular}{lrrrrrr}
\hline
&$\phi_1$&$\phi_2$&$\phi_3$&$\phi_4$&$\phi_5$&$\phi_{j\ge6}$\\
\hline
\ref{fig:fBO9}(a)&\tdeg{162.4}&\tdeg{162.4}&\tdeg{313.4}&\tdeg{284.5}&\tdeg{0.0}&A\\
\ref{fig:fBO9}(b)&\tdeg{46.6}&\tdeg{255.5}&\tdeg{255.5}&\tdeg{75.5}&\tdeg{0.0}&A\\
\ref{fig:fBO9}(c)&\tdeg{282.1}&\tdeg{339.5}&\tdeg{339.4}&\tdeg{159.4}&\tdeg{114.6}&S\\
\hline
\end{tabular}
\caption{Explicit pulse phases for some composite pulses comprising nine $\pi$ pulses and depicted in Fig.~\ref{fig:fBO9}; phases $\phi_{j\ge6}$ can be obtained by noting that each pulse is either antisymmetric (A) or symmetric (S).}\label{tab:phi9}
\end{table}

Next I consider the design of symmetric pulse sequences.  Solving the constraint equations is quite complex in this case, but the solution is found to be
\begin{equation}
\mbox{\boldmath$(\phi)$}
=(\alpha,\,\beta,\,\beta,\,\beta-\pi,\,2\beta-2\alpha,\,\beta-\pi,\,\beta,\,\beta,\,\alpha)
\end{equation}
with
\begin{equation}
\beta=2\alpha+\arccos[-(1+2\cos\alpha)/2]
\end{equation}
and
\begin{equation}
\alpha=-\arccos[(4-\sqrt{10})/4].
\end{equation}
The resulting phases are listed in Table~\ref{tab:phi9}, and the performance of this composite pulse is shown in Fig.~\ref{fig:fBO9}(c).

\section{Other approaches}
It is useful to compare these results with other approaches that have been used to develop composite pulses with simultaneous error tolerance.  An early example is a sequence of nine $\pi$ pulses described by Alway and Jones \cite{Alway2007} with phases
\begin{equation}
\mbox{\boldmath$(\phi)$}
=\left(0,\,\psi,\,3\psi,\,3\psi,\,\psi,\,\pi-\psi,\,-\psi,\,\pi+\psi,\,\psi\right)
\end{equation}
where $\psi=\arccos(-1/4)$ as usual.  This composite pulse suppresses first and second order error terms for pulse strength errors and first order off-resonance errors, but this can be achieved with a five pulse sequence as discussed above.  It should, however, be noted that this pulse was developed as an illustration of a general approach to designing composite pulses, in which individual errors are removed one by one \cite{Brown2004,Brown2005,Alway2007}, rather than being optimized for these particular properties.

A more interesting approach is the family of concatenated composite pules developed by Bando, Ichikawa, Kondo and Nakahara \cite{Ichikawa2011,Bando2013}, which seek to achieve simultaneous error correction by nesting correction of off-resonance errors within correction of pulse strength errors, or \textit{vice versa}.  Once again simultaneous error tolerance can be achieved, but the results are no better than for sequences of five $\pi$ pulses.  However these concatenated pulse can be extended to pulses with rotation angles other than $\pi$, and here they may be more useful.  A similar approach has been explored by Merrill and Brown \cite{Merrill2012}.

If, however, simultaneous error tolerance is not required then previous approaches to composite pulse design are likely to be preferable in many cases.  In particular for the pulse strength errors the fully time symmetric BB1 sequence suppresses first and second order errors with a total sequence length equivalent to five $\pi$ pulses, and at no cost in increased sensitivity to off-resonance errors.  If even more effective error suppression is required then symmetrised $\textrm{W}_n$ pulses \cite{Husain2013} can be used.  Similarly, for off-resonance errors the CORPSE pulse sequence \cite{Cummins2000,Cummins2003} can be used to suppress the first order error and almost (but not entirely) remove the second order error, with a total sequence length equivalent to between four and five $\pi$ pulses, and at no cost in increased sensitivity to pulse strength errors.  All of these composite pulses have the advantage that they can be generalized to other rotation angles.

\section{Conclusions}
It is clear that it is possible to design quite short composite \NOT\ gates with excellent simultaneous tolerance of pulse strength and off-resonance errors: comparing Fig.~\ref{fig:fBO9}(c) with Fig.~\ref{fig:fNOB}(a) shows that the sequence of nine $\pi$ pulses has a similar region of parameter space inside the $10^{-6}$ contour as the naive single $\pi$ pulse has inside the contour at $10^{-2}$.

The insight from NMR studies that it is useful to concentrate on the geometric form of the error, rather than proceeding blindly with algebraic minimisation, is certainly correct for pulses up to $n=9$.  Beyond this it is less clear how such geometric constraints should be applied, as suitable closed equilateral polygons can be formed in many ways.  However it is less obvious that the second NMR insight, that antisymmetric composite pulses which suppress first order errors also automatically suppress second order errors, is as useful, and for many purposes in QIP symmetric composite pulses, designed by combining geometric and algebraic elimination of error terms, are likely to be more appropriate.

One further advantage of antisymmetric pulses designed to tackle pulse strength errors is that they can be iteratively nested \cite{Husain2013} to produce composite pulses with exceptionally broad error tolerance.  This approach also works with pulses designed for simultaneous error tolerance; in particular ASBO-9 composite pulses can be nested to produce a composite pulse which removes both error terms up to ninth order, so that the infidelity is eighteenth order in both $\epsilon$ and $f$.  This behaviour can ultimately be traced to the fact that these antisymmetric pulses can suppress multiple even order error terms \cite{Merrill2012}.  However, nested ASBO-9 pulses comprise 81 separate $\pi$ pulses, and such long composite pulses are not considered here.


\begin{acknowledgments}
I thank Ron Daniel for helpful conversations.
\end{acknowledgments}

\bibliography{../all}
\end{document}